\documentclass[aps,preprint,epsfig,rotate]{revtex4}
\begin{document}
\title{On the interpolation formula for the bound state energies of atomic systems}

 \author{Alexei M. Frolov}
 \email[E--mail address: ]{afrolov@uwo.ca}

\affiliation{Department of Applied Mathematics \\
 University of Western Ontario, London, Ontario N6H 5B7, Canada}

\date{\today}

\begin{abstract}

By using results of highly accurate computations of the total energies of a large number of few-electron atoms we construct a few interpolation formulas 
which can be used to approximate the total energies of bound atomic states. In our procedure the total energies of atomic states $E$ are represented as
a function of the electric charge of atomic nucleus $Q$ and the total number of bound electrons $N_e$. Some general properties of the $E(Q, N_e)$ function 
are investigated. The knowledge of the $E(Q, N_e)$ function allows one to determine the total (and binding) energies of these states in arbitrary atoms and 
ions with different $Q$ and $N_e$. 

\noindent 
PACS number(s): 31.10.+z and 31.15.-p and 31.90.+s


\end{abstract}

\maketitle
\newpage

In this short communication we discuss accurate and relatively simple interpolation formulas which can be used to predict the total energies and other 
bound state properties in various atoms and ions. Recently, a very substantial progress has been achieved in highly accurate computations of the bound states in 
few-electron atoms and ions. Based on the results of such calculations we can construct a number of different interpolation formulas for the total and binding 
energies of atomic few-electron systems, i.e. for atomic systems with different nuclear charges $Q$ and different number of bounded electrons $N_e$. Analogous
formulas can be constructed for other bound state properties, including inter-particle delta-functions, various single-, two- and three-particles properties and
properties which are determined by the expectation values of some singular operators. In this study we shall not discuss interpolation formulas for arbitrary bound 
state properties. Furthermore, below we restrict ourselves to the consideration of the total energies of the bound states in the non-relativistic atoms and ions.      
First, note that the overall accuracy of the $Q^{-1}$ expansions constructed for the ground states in the two-, three- and four-electron atomic systems can be 
considered as outstanding. In this study we want to make the following step and develop (and later apply) the `universal formula' proposed in \cite{FrWa2008}, which 
allows one to predict the total energies of arbitrary atomic systems with the known values of $Q$ and $N_e$ to very high numerical accuracy and without actual atomic 
calculations. In other words, we need to construct the universal function $E(Q, N_e)$ of the two (integer) arguments and investigate its properties.   

At the first step of our analysis we need to determine the total energies $E$ and non-relativistic wave functions as the solutions of the Schr\"{o}dinger 
equation \cite{FrFi} for the bound states $H \Psi = E \Psi$, where $E < 0$ and $H$ is the non-relativistic Hamiltonian of the $N_e-$electron ion/atom 
\begin{equation}
 H = -\frac{\hbar^{2}}{2 m_{e}} \sum_{i=1}^{N_e} \nabla^{2}_{i} - Q e^2 \sum_{i=1}^{N_e} \frac{1}{r_{in}} + e^2 \sum^{N_e-1}_{j=1} 
 \sum^{N_e}_{i=2 (i > j)} \frac{1}{r_{ij}} \label{Ham}
\end{equation}
where $\nabla_{i} = \Bigl( \frac{\partial}{\partial x_{i}}, \frac{\partial}{\partial y_{i}}, \frac{\partial}{\partial z_{i}} \Bigr)$, $i = 1, 2, \ldots, N_e$ and 
the notation $n$ stands for the atomic nucleus. In Eq.(\ref{Ham}) the notation $\hbar$ designates for the reduced Planck constant, i.e. $\hbar = \frac{h}{2 \pi}$, 
and $e$ is the elementary electric charge. In this study all masses of the atomic nuclei are assumed to be infinite. In general, it is very convenient to perform all 
bound state calculations in atomic units where $\hbar = 1, m_e = 1$ and $e = 1$. In these units the velocity of light in vacuum $c$ numerically coincides with the 
inverse value of the dimensionless fine structure constant, i.e. $c = \alpha^{-1}$, where $\alpha = \frac{e^2}{\hbar c} \approx$ 7.2973525698$\cdot 10^{-3}$. In 
atomic units the Hamiltonian, Eq.(\ref{Ham}), is written in the form
\begin{equation}
 H = -\frac12 \sum_{i=1}^{N_e} \nabla^{2}_{i} - Q \sum_{i=1}^{N_e} \frac{1}{r_{in}} + \sum^{N_e-1}_{i=1} \sum^{N_e}_{i=2 (i > j)} 
 \frac{1}{r_{ij}} \label{Ham1}
\end{equation}

As follows from Eq.(\ref{Ham1}) the Hamiltonian $H$ is a continuous operator-function of the nuclear charge $Q$, or in other words, the nuclear charge $Q$ is the continuous 
parameter (or control parameter) of this Hamiltonian. By applying the Poincare theorem one finds that all eigenvalues of the Hamiltonian $H$, Eq.(\ref{Ham1}), and all its 
eigenfunctions are the continuous functions of $Q$. In applications to real atoms the nuclear charge $Q$ expressed in atomic units is always an integer number. Let us assume 
that we know the actual atomic wave function $\Psi$ (or $\mid \Psi \rangle$). Then we can reduce Eq.(\ref{Ham}) by the following equation for the three expectation values
\begin{equation}
 E = \langle H \rangle = -\frac12 N_e \langle \nabla^{2}_{1} \rangle - N_e Q \langle \frac{1}{r_{1n}} \rangle + \frac{N_e (N_e - 1)}{2} 
 \langle \frac{1}{r_{12}} \rangle \label{Ham2}
\end{equation}
where $E$ is the total energy of the bound state, while $\langle \nabla^{2}_{1} \rangle, \langle \frac{1}{r_{1n}} \rangle = \langle r^{-1}_{1n} \rangle, \langle \frac{1}{r_{12}} 
\rangle = \langle r^{-1}_{12} \rangle$ are the expectation values of the electron kinetic energy, electron-nucleus (attractive) potential energy and electron-electron repulsion, 
respectively. In derivation of Eq.(\ref{Ham2}) the fact that all atomic electrons identical particles. As it can be seen from Eq.(\ref{Ham2}) the total energy $E$ is a 
function of the two parameters $Q$ and $N_e$ which are both integer. Formally, to determine the exact (or analytical) form of the $E(Q, N_e)$ function we need to determine 
three expectation values mentioned in Eq.(\ref{Ham2}), i.e. the $\langle \nabla^{2}_{1} \rangle = - \langle {\bf p}^{2}_{1} \rangle, \langle r^{-1}_{1n} \rangle$ and $\langle 
r^{-1}_{12} \rangle$. In reality, for Coulomb systems one also finds an additional condition which is widely known as the `virial theorem'. The virial is written in the form $2 
\langle T \rangle = - \langle V \rangle$, where $T$ is the operator of kinetic energy, while $V$ is the operator of the potential energy. The explicit forms of these operators 
are:
\begin{equation}
 T = -\frac12 \sum_{i=1}^{N_e} \nabla^{2}_{i} \; \; \; and \; \; \;  
 V = - Q \sum_{i=1}^{N_e} \frac{1}{r_{in}} + \sum^{N_e-1}_{i=1} \sum^{N_e}_{i=2 (i > j)} \frac{1}{r_{ij}} \label{TV}
\end{equation}
Note that the virial theorem can be written in one of the following forms: $E = -\langle T \rangle, \frac12 \langle V \rangle = E$, etc. In general, by applying the virial 
theorem we can reduce the total number of `unknown' expectation values in the right-hand side of Eq.(\ref{Ham2}) from three to two. For instance, the expression of the total 
energy $E$ in terms of $\langle \frac{1}{r_{1n}} \rangle$ and $\langle \frac{1}{r_{12}} \rangle$ expectation values is
\begin{equation}
 E = \langle H \rangle = -\frac12 N_e Q \langle \frac{1}{r_{1n}} \rangle + \frac{N_e (N_e - 1)}{4} \langle \frac{1}{r_{12}} \rangle \label{Energy}
\end{equation}

As follows from Eq.(\ref{Energy}) to obtain the explicit formula for the $E(Q, N_e)$ function we need to derive analogous formulas for the $\langle \frac{1}{r_{en}} \rangle$ and 
$\langle \frac{1}{r_{ee}} \rangle$ expectation values. Briefly, this means that these expectation values must be expressed as explicit functions of $Q$ and $N_e$. In reality, it is 
not possible to derive any closed analytical expression for the $\langle \frac{1}{r_{en}} \rangle$ and/or $\langle \frac{1}{r_{ee}} \rangle$ expectation values written as a 
function of $Q$ and $N_e$. An obvious exclusion is the Thomas-Fermi method \cite{TomFer}. By  using other methods which are more accurate than Thomas-Fermi method it is impossible 
to derive the closed analytical formulas for the $\langle \frac{1}{r_{en}} \rangle$ and $\langle \frac{1}{r_{ee}} \rangle$ expectation values. This means that it is impossible to 
obtain any closed analytical formula for the $E(Q, N_e)$ function. However, the `atomic function' $E(Q, N_e)$ can be approximated to very good accuracy by using results of highly 
accurate numerical calculations for a large number of a few-electron atoms/ions. Formally, if the total number of bounded electrons $N_e$ is fixed, then we are dealing with the 
so-called $Q^{-1}$ expansions for the total energies of a number of atoms/ions with different $Q$. For two-electron atoms and ions such series are well known since the middle of 
1930's and first papers by Hylleraas for two-electron ions (see, e.g., \cite{BS} -\cite{Eps} and references therein). Analogous series for three- and four-electron atomic systems 
were not used in applications, since they provided a very modest overall accuracy which was not sufficient for accurate evaluations. 

At this moment the situation with accurate numerical calculations of the three- and four-electron atoms and ions has changed. Currently, we have a large number of highly accurate 
results for the two-electron ions and quite a few different sets of accurate numerical results obtained for three- and four-electron atoms and ions (see, e.g., \cite{Yan}, \cite{Sims} 
and references therein). These results allow us to construct some accurate $Q^{-1}$ expansions which also describe the total energies of the three- and four-electron atomic systems. 
Based on these formulas for the $Q^{-1}$ expansions for two-, three- and four-electron atoms and ions we can try to guess the explicit formulas for an universal function $E(Q, N_e)$, 
where $N_e$ is the total number of bounded electrons. This problem has a fundamental value for whole atomic physics. Indeed, if we know the $E(Q, N_e)$ function, then we can predict 
the total energy of an arbitrary atom/ion with the given $Q$ and $N_e$ to high accuracy which is sufficient for many actual problems known from stellar astrophysics, physics of 
high-temperature plasmas, etc. In reality, accurate predictions of the total and binding energies of atoms/ions with different $Q$ and $N_e$ was an ultimate goal for many generations 
of atomic physicists. In this study we show that currently we are very close to fulfill this goal.       
   
Accurate non-relativistic energies $E$ of a large number of the ground states in the two-, three- and four-electron atoms/ions can be found in Table I. All total energies are given in 
Table I in atomic units. The total energies of the two-electron (or helium-like) atoms and ions from Table I have been determined to very high accuracy with the use of our computational 
procedure which allows one to perform calculations with 3500 exponential basis functions in the wave functions. All basis functions are written in the relative coordinates $r_{32}, r_{31}$
and $r_{21}$. For three-electron (or lithium-like) ions (and Li atom) such energies have been taken from \cite{Yan}, while for four-electron atoms/ions they were chosen from \cite{Sims}. 
The overall accuracy of the ground state energies in three- and four-electron atomic systems is still significantly lower than the analogous accuracy for two-electron atoms and ions (see 
Table I). Note also that the Li$^{-}$ ion, i.e. atomic system with $Q = 3$ and $N_e = 4$, is bound, i.e. its ground $2^1S$-state is stable, but our current variational results for this 
system (e.g.,$E$ = -7.5007185 $a.u.$) are not highly accurate. This is the reason why we exclude this ion from Table I. By using the total energies of all atoms/ions mentioned in Table I 
(with the same value of $N_e$) one can determine some numerical coefficients in the $Q^{-1}$ expansion
\begin{equation}
 E(Q) = a_2 Q^2 + a_1 Q + a_0 + b_1 Q^{-1} +  b_1 Q^{-1} + b_2 Q^{-2} + b_3 Q^{-3} + \ldots \; \; \; \label{Qexp}
\end{equation}
where $a_2, a_1, a_0$ are the coefficients of the regular part of the Laurent expansion (or series), while $b_1, b_2, \ldots$ are the coefficients of the principal part of the Laurent series 
$E(Q)$, Eq.(\ref{Qexp}). Note that the $Q^{-1}$-expansion (or $Q^{-1}$-series), Eq.(\ref{Qexp}), is a typical `asymptotic expansion'. Briefly, this means that after some $n \ge n_{max}$ all 
coefficients $b_{n}$ in Eq.(\ref{Qexp}) rapidly increase with $n$. Contributions of the corresponding terms also rapidly increase with $n$ and the total sum computed with the use of 
Eq.(\ref{Qexp}), which includes such `growing terms', has nothing to do with the original problem. Briefly, this means that in such cases the $Q^{-1}$ expansion cannot be used to approximate 
the actual total energies. To avoid this problem we need to restrict the total number of terms in Eq.(\ref{Qexp}). For instance, if we use $N = 18$ values of the total energies $E(Q)$ 
(computed for eighteen different values of $Q$), then the total number of terms in Eq.(\ref{Qexp}) can be 8, or 10, but not 16, or 18. An universal criterion can be formulated in the following 
form: overall contribution to the total sum, Eq.(\ref{Qexp}), from the last term must be smaller (and even much smaller) than analogous contribution from the pre-last term. 

First ten coefficients of the $Q^{-1}$ expansion, Eq.(\ref{Qexp}), determined from the results of highly accurate numerical calculations of few-electron atoms and ions mentioned in Table I can 
be found in Table II. These coefficients can be used in applications of the $Q^{-1}$ expansion, Eq.(\ref{Qexp}), to other two-, three- and four-electron atoms and ions. The overall accuracy of 
the $Q^{-1}$ expansion, Eq.(\ref{Qexp}), for the total energies of the ground states in these atoms and ions is outstanding and can be evaluated as $1 \cdot 10^{-9} - 5 \cdot 10^{-12}$ $a.u.$ 
of the total energies. In general, the $Q^{-1}$ expansion, Eq.(\ref{Qexp}), can be used in applications to different iso-electron atomic systems. However, due to numerous problems in accurate 
computations of the three-, four- and many-electron atoms and ions the most successful applications of the $Q^{-1}$ expansion are still restricted to the two-electron (or helium-like) atoms and 
ions. It should be mentioned that similar $Q^{-1}$ expansions can be used for other bound state properties, e.g., to predict interparticle distances, expectation values of some delta-functions, 
etc. As mentioned above in this study we restrict ourselves to the total energies only and (brief discussion of the interpolation formulas for other bound state properties can be found, e.g., in 
\cite{FrWa2008}).

Based on the results from Tables I and II we can make the new step which must lead to better understanding of the structure of bound state spectra in atoms and ions. Instead of dealing with many 
different $E(Q)$ functions constructed for each series of iso-electron atomic systems, i.e. atoms/ions with the same $N_e$, we introduce a `universal function' $E(Q, N_e)$ which depends upon two 
integer numbers: $Q$ (nuclear electric charge) and $N_e$ (total number of bound electrons). If $N_e$ is fixed, e.g., $N_e = 2$, then the corresponding function $E(Q, N_e = 2) = E(Q, 2)$ must 
coincide with the function $E(Q)$ known for the two-electron atomic systems. There are few possible approaches which can be used to determine the universal $E(Q, N_e)$ function. In this study we 
apply the so-called direct approach which is based on the results presented in Table II. In fact, we shall assume below that coefficients presented in all columns of Table II correspond to one 
`universal' function $E(Q, N_e)$. Numerical differences in these coefficients can only be related with the variations in $N_e$. For instance, consider the first coefficients $a_2$ from Table II. 
The exact value of these coefficients are $-1, -\frac98$ and $-\frac54$ for $N_e$ = 2, 3, and 4, respectively. Therefore, the following general formula can be written in the form
\begin{equation}
 a_2 = -\Bigl( \frac{8 + N_e - 2}{8} \Bigr) = -\Bigl( \frac{6 + N_e}{8} \Bigr) \; \; \; \label{coefa2}
\end{equation}      
where $N_e$ is the total number of bounded electrons in the atom/ion. As follows from Table II the analogous expression for the second coefficient can written in 
the following form 
\begin{equation}
 a_1 = \frac58 + a^{(1)}_1 (N_e - 2) + a^{(2)}_1 (N_e - 2)^{2} \; \; \; \label{coefa1}
\end{equation}
where $a^{(1)}_1$ and $a^{(2)}_1$ are the two unknown coefficients which are determined with the use of numerical values for this coefficient from the second and third columns of 
Table II. The formula, Eq.(\ref{coefa1}), can be re-written into a slightly different form 
\begin{equation}
 a^{(N_e)}_1 = \frac58 + \Bigl[ a^{(3)}_1 - \frac58 \Bigr] (N_e - 2) + \Bigl[ a^{(4)}_1 - 2 a^{(3)}_1 + \frac58 \Bigr] \frac{(N_e - 2) (N_e - 3)}{2} \; \; \; \label{coefa2a}
\end{equation}
where $a^{(3)}_1$ and $a^{(4)}_1$ are the coefficients from Table I for the three- and four-electron atomic system, respectively. The unknown value of the $a_1$ coefficient for atomic system with 
$N_e-$electrons is designated in the left-hand side of Eq.(\ref{coefa2a}) as $a^{(N_e)}_1$. The formula, Eq.(\ref{coefa2a}), can be generalized to more complex cases, e.g., 
\begin{equation}
 a^{(N_e)}_0 = a^{(2)}_0 + \Bigl[ a^{(3)}_0 - a^{(2)}_0 \Bigr] (N_e - 2) + \Bigl[ a^{(4)}_0 - 2 a^{(3)}_0 + a^{(2)}_0  \Bigr] \frac{(N_e - 2) (N_e - 3)}{2} \; \; \; \label{coefa3}
\end{equation}
where $a^{(N_e)}_0$ is the $a_0$ coefficient from Eq.(\ref{Qexp}) defined for atomic systems with $N_e$ bound electrons. Analogously, for $b_{k}$ coefficients from Eq.(\ref{Qexp}) one
finds
\begin{equation}
 b^{(N_e)}_k = b^{(2)}_k + \Bigl[ b^{(3)}_k - b^{(2)}_k \Bigr] (N_e - 2) + \Bigl[ b^{(4)}_k - 2 b^{(3)}_k + b^{(2)}_k \Bigr] \frac{(N_e - 2) (N_e - 3)}{2} \; \; \; \label{coefa3a}
\end{equation}
where notation $a^{(N_e)}_k$ stands for the $b_k$ coefficients ($k = 1, 2, 3, \ldots$) from Eq.(\ref{Qexp}) defined for atomic systems with $N_e$ bound electrons, where $N_e = 2, 3, 4, \ldots$. 
This expression is, in fact, the Taylor-Maclaurin expansion for the $b^{(N_e)}_k$ coefficients upon the total number of bound electrons $N_e$. Note that the formulas Eqs.(\ref{coefa2}) - 
(\ref{coefa3a}) for the coefficient $a_2 - a_0$ and $b_k$ ($k = 1, 2, \ldots$) are essentially exact upon $N_e$ for all two-, three- and four-atoms/ions presented in Table I. This means that the 
total energies of these ions are reproduced with the same accuracy which is provided by their $Q^{-1}$ expansion.

As follows from Eq.(\ref{coefa2}) the coefficient $a_2$ is the ratio of two small integer numbers, and numerator of this fraction is a linear function of $N_e$. This follows from the fact that the 
main contribution to the total energy of any atom/ion comes from the $N_e Q^{2}$-term which represents the leading term in the electron-nucleus attraction. In atomic units this term equals to an 
integer, or semi-integer number (for an arbitrary atom/ion). The coefficient $a_1$ for two-electron atomic systems is also a simple fraction, e.g., $\frac58$ for the ground states. The same conclusion 
is true for other bound states, e.g., for the $2^3S-$triplet states, in the two-electron atoms/ions \cite{Fro2005}. 

In general, the analysis of the $E(Q, N_e)$ function(s) is very similar to operations with the Weiz\"{a}cker mass formula in nuclear physics. Such a similarity follows from the fact that 
in both cases we are dealing with the finite Fermi systems \cite{Migdal}. Note that atoms and ions were considered as the finite Fermi systems in atomic physics since first papers published in earlier 
1960's \cite{Shull} and \cite{Scherr} (see also \cite{March} and references therein). More recent references can be found, e.g., in \cite{Chak}. This general theory allows one to obtain the asymptotic 
form of the $E(Q, N_e)$ function at very large $Q$ and $N_e$ \cite{March}. A few other advantages of that theory include the correct theoretical expressions for the $a_0$ and $a_1$ coefficients (see, 
Eqs.(\ref{coefa2}) - (\ref{coefa1})) which either coincide with ours (see Table II), or very close to them. The coefficients $a_2, b_1, b_2, \ldots$ predicted in the general theory are known only 
approximately, even in the case of the ground (bound) atomic states. For excited atomic states the accuracy of predictions of this theory is substantilally lower. In many cases applications of the 
general theory to the excited atomic states \cite{March} lead to very inaccurate predictions of ionization potentials and other properties. Such a situation with excited states lead to a conclusion that 
applications of the general theory to the ground states in atomic systems only. However, in this case we have to face the following crucial question: {\it why do we need to perform dozens of highly 
accurate, atomic computations for construction of the accurate interpolation formula for the total energies, if in modern atomic physics highly accurate computations of any ground state in arbitrary 
atom/ion are significantly faster}? It is clear that if we cannot generalize our interpolation formulas to the excited atomic states, then chances of the `general theory' \cite{Shull} - \cite{March} to 
survive in the future are very low.   

Our approach based on the use of highly accurate computational data for different atoms/ions allows one to reconstruct the $E(Q, N_e)$ function can be applied, in principle to any atomic state,
including excited states. As is well known from atomic spectroscopy, the bound state spectrum of any multi-electron atom/ion is represented as a set of different terms, where each term has its unique 
quantum numbers of angular momentum $L$ and total electron spin $S$. For our analysis this means that the function $E(Q, N_e)$ (total energy) defined above must be labeled by the two indexes $L$ and 
$S$, which are good (= conserving) quantum numbers for an isolated atom/ion with $N_e$ bound electrons. These two quantum numbers are the labels of the corresponding atomic term. The bound state spectrum
of any atom is represented as a combination of different $LS$-terms. An ultimate goal is to approximate the total (non-relativistic) energies of all bound states from all possible terms which can be 
found  in real atoms and ions. In the `general theory' \cite{Shull} and \cite{Scherr} this problem is extremely complex, since often the atomic $LS-$term does not exist in atoms/ions with fewer electrons. 
Formally, this means that we need to construct the $Q^{-1}$ expansions for each different atomic $LS-$term. 

In part, such a strategy already works for the ground atomic states. Indeed, the ground states in different atoms correspond to the different terms, e.g., the ground state in the B-atom (five bound 
electrons) is the $2^{1}P$-state, analogous state in the carbon atom is the $2^{2}P$-state. It is clear that by including future highly results for atoms/ions with larger number of electrons we can reach 
the ground states of the Fe atom (${}^{5}D_{4}$-term), Co atom (${}^{4}D_{\frac92}$-term), etc. However, if it is possible to connect all ground atomic states by one interpolation formula, then we can do 
the same for the excited atomic states too. Let us describe our current vision of this two-stage procedure. At the first stage the approach is based on the formulas, Eqs.(\ref{coefa2}), (\ref{coefa2a}), 
(\ref{coefa3}) and (\ref{coefa3a}) allows one to determine other unknown coefficients $a_0, a_1, a_2, b_1, b_2, \ldots$. The arising formulas are simple and convenient in applications to atoms and ions. 
The known coefficients $a_0, a_1, a_2, b_1, b_2, \ldots$ are used in the `usual' $Q^{-1}$-expansion  
\begin{eqnarray}
 E(Q, N_e) = a_2(N_e) Q^2 &+& a_1(N_e) Q + a_0(N_e) + b_1(N_e) Q^{-1} + b_1(N_e) Q^{-1} \nonumber \\
 &+& b_2(N_e) Q^{-2} + b_3(N_e) Q^{-3} + \ldots \; \; \; \label{Qexp2}
\end{eqnarray}
where now all coefficients are the functions of the total number of bound electrons $N_e$. In this study all coefficients in Eq.(\ref{Qexp2}) are constructed as polynomial functions of $N_e$. The 
$Q^{-1}$-expansion, Eq.(\ref{Qexp2}), provides high numerical accuracy for the total energies of all two-, three- and four-electron atomic systems presented in Table I. Very likely, that the 
analytical expression for each of the $b_i(N_e)$ coefficients ($i$ = 1, 2, $\ldots$) in Eq.(\ref{Qexp2}) is more complicated than a simple polynomial in $N_e$. It is clear that such an expansion 
must also include the negative powers of $N_e$. In reality, we can investigate the $E(Q, N_e)$ function in detail when we can obtain highly accurate results for five- and six-electron atoms and 
ions. When highly accurate computations of the five- and six-electron atoms will be completed, then we can substantially improve our current knowledge of the $E(Q, N_e)$ function(s). Moreover, we 
can derive some compact and accurate formulas for numerical approximations of these functions for different bound atomic states. This is an answer to an old question about possibility to find an 
analytical formula for the total non-relativistic energies of bound states in atomic system which contains $N_e$ bound electrons moving in the field of infinitely heavy nucleus with the electric 
charge $Q e$. In this study to approximate the $E(Q, N_e)$ function we have restricted to the ten-term formula for the $Q^{-1}$ expansion, Eq.(\ref{Qexp}). In part, such a restriction is related with 
relatively low accuracy of the computational energies obtained for the three- and four-electron atoms/ions.  

At the second stage of the procedure we need to find relations between coefficients of these series, define these coefficients as the functions of $N_e$, etc. As mentioned above some bound $LS-$state (or 
$LS$-term) may not exist for atoms/ions with fewer electrons. Therefore, we need to solve the problem of genealogical relation between terms (or bound sates) in atoms/ions with different number(s) of 
bound electrons $N_e$. To explain this problem let us assume that we have determined known function $E_{LS}(Q, N_e)$ (total energy) for atoms/ions with the same number bound electrons $N_e$ ($Q$ is varied). 
Now, suppose that the total number of bound electrons increases by one, i.e. $N_e \rightarrow N_e + 1$. Now, we have the new term $L^{\prime}S^{\prime}$ and the new function $E_{L^{\prime}S^{\prime}}(Q, 
N_e+1)$ (total energy). We need to predict possible numerical values of $L^{\prime}$ and $S^{\prime}$ quantum numbers. As it follows from the fundamental principles of atomic theory for the new spin quantum 
number we have $S^{\prime} = S \pm \frac12$, if $S \ne 0$, and $S^{\prime} = S + \frac12$, if $S = 0$. To predict the new value of $L^{\prime}$ we have to know the angular momentum $\ell$ of the additional 
electron. If we know this value, then one finds that all possible $L^{\prime}$ values are located between the two following limits $\mid L - \ell \mid$ (lower limit) and $L + \ell$ (upper limit), i.e. $\mid 
L - \ell \mid \le L^{\prime} \le L + \ell$. If these conditions for the $L^{\prime}$ and $S^{\prime}$ are obeyed, then the total energies $E_{LS}(Q, N_e)$ and $E_{L^{\prime}S^{\prime}}(Q, N_e+1)$ can be 
used in one series. 

As follows from the results of our study the function $E(Q, N_e)$ can be constructed for all bound states in multi-electron atomic systems, if (and only if) their terms are directly connected by the 
`selection' rule mentioned above. For instance, in this study we discuss the total energies of two-, three- and four-electron atoms/ions which have their ground $1^{1}S-, 2^{2}S-$ and 
$2^{1}S-$states, respectively. These bound states have different multiplicities, i.e. they are singlets and doublets states. Suppose we want to include in our analysis five-electron atoms/ions (B-like 
ions) for which the ground state is the $2^{2}P-$state. The `selection rules' mentioned above works in this case. Therefore, we can construct the $E(Q, N_e)$ function for the $1^{1}S-, 2^{2}S-, 
2^{1}S-$ and $2^{2}P-$states in the two-, three-, four- and five-electron atoms/ions, respectively. However, if we want to evaluate, e.g., the total energies of the $3^{4}D-$states in the five-electron 
ions, then we have to use a slightly different consequence of bound states the two-, three- and four-electron atoms/ions, respectively. A natural choice in this case is to consider the bound $1^{1}S-, 
2^{2}S-$ and $2^{3}P-$states in the two-, three- and four-electron atoms/ions, respectively. The total energies determined for the bound $3^{4}D-$states in the five-electron atoms/ions perfectly complete
the data (total energies) computed for the bound $1^{1}S-, 2^{2}S-$ and $2^{3}P-$states in the two-, three- and four-electron atoms/ions. On the other hand, it is clear that our total energies from the 
third column of Table I (total energies of the $2^{1}S-$states in the four-electron atoms/ions) are useless in this case. In other words, these energies cannot be used to predict the total energies of the 
bound $3^{4}D-$states in five-electron atoms/ions. Formally, this means that to construct highly accurate interpolation formula for the total energies of the bound $3^{4}D-$states in the five-electron 
atoms and ions one needs to perform a number of separate, highly accurate computations of the bound $2^{3}P-$states in the four-electron atomic systems. In general, the total energies of any bound state
in multi-electron atomic system can be predicted to relatively high accuracy, if we know a number of total energies of bound states in atomic systems with fewer electrons. SUch bound states in atomic 
systems with fewer electrons correspond to the different $LS-$terms. However, all these bound states (or terms) must be related to each other by the selection rules mentioned above. An additional problem
follows form the fact that some of the atomic terms can be connected by using different atomic terms in the systems with fewer bounded electrons. It is clear that the procedure must be self-correlated. This 
means that we must have some additional relations between the $E_{LS}(Q, N_e)$ and $E_{L^{\prime}S^{\prime}}(Q, N^{\prime}_e)$ functions constructed for different atomic terms and for atoms/ions which contain 
different number(s) of bound electrons. This interesting question cannot be answered at the current level of theoretical and computational development. 

\begin{center}
 {\Large Appendix} 
\end{center}

The formulas Eqs.(\ref{coefa2}), (\ref{coefa2a}), (\ref{coefa3}) and (\ref{coefa3a}) for the coefficients $a_2, a_1, a_0$ and $b_1, b_2, \ldots$ from the main text can be re-written in a number of 
different forms which are often more convenient in applications. In this Appendix we present the explicit formulas for the coefficients $a_2, a_1, a_0$ and $b_1, b_2, \ldots$ for five- and six-electron 
atomic systems (i.e. toms and ions). In all these cases the formula for the $a_2$ coefficient coincides with Eq.(\ref{coefa2}) (here we do not want to repeat it). The formulas for other coefficients from 
Eq.(\ref{Qexp2}) take the following forms. For $N_e = 5$ one finds:
\begin{eqnarray}
 a^{(N_e=5)}_1 &=& \frac58 + \Bigl[ a^{(3)}_1 - \frac58 \Bigr] (N_e - 2) + \Bigl[ a^{(4)}_1 - 2 a^{(3)}_1 + \frac58 \Bigr] \frac{(N_e - 2) (N_e - 3)}{2} \nonumber \\ 
 &+& \Bigl[ a^{(5)}_1 - 3 a^{(4)}_1 + 3 a^{(3)}_1 
 - \frac58 \Bigr] \frac{(N_e - 2) (N_e - 3) (N_e - 4)}{6} \; \; \; \label{coefa2c} \\
 a^{(N_e=5)}_0 &=& a^{(2)}_0 + \Bigl[ a^{(3)}_0 - a^{(2)}_0 \Bigr] (N_e - 2) + \Bigl[ a^{(4)}_0 - 2 a^{(3)}_0 + a^{(2)}_0  \Bigr] \frac{(N_e - 2) (N_e - 3)}{2} \nonumber \\
 &+& \Bigl[ a^{(5)}_2 - 3 a^{(4)}_2 + 3 a^{(3)}_2 - a^{(2)}_2 \Bigr] \frac{(N_e - 2) (N_e - 3) (N_e - 4)}{6} \; \; \; \label{coefa3c}
\end{eqnarray}
and
\begin{eqnarray}
 b^{(N_e=5)}_k &=& b^{(2)}_k + \Bigl[ b^{(3)}_k - b^{(2)}_k \Bigr] (N_e - 2) + \Bigl[ b^{(4)}_k - 2 b^{(3)}_k + b^{(2)}_k \Bigr] \frac{(N_e - 2) (N_e - 3)}{2} \nonumber \\
 &+& \Bigl[ b^{(5)}_k - 3 b^{(4)}_k + 3 b^{(3)}_k - b^{(2)}_k \Bigr] \frac{(N_e - 2) (N_e - 3) (N_e - 4)}{6} \; \; \; \label{coefb3c}
\end{eqnarray}
where $N_e = 5$ in the right-hand sides of these equations. 

Analogous formulas for the $N_e = 6$ are 
\begin{eqnarray}
 a^{(N_e=6)}_1 &=& \frac58 + \Bigl[ a^{(3)}_1 - \frac58 \Bigr] (N_e - 2) + \Bigl[ a^{(4)}_1 - 2 a^{(3)}_1 + \frac58 \Bigr] \frac{(N_e - 2) (N_e - 3)}{2} \nonumber \\ 
 &+& \Bigl[ a^{(5)}_1 - 3 a^{(4)}_1 + 3 a^{(3)}_1 - \frac58 \Bigr] \frac{(N_e - 2) (N_e - 3) (N_e - 4)}{6} \; \; \; \label{coefa2d} \\
 &+& \Bigl[ a^{(6)}_1 - 4 a^{(5)}_1 + 6 a^{(4)}_1 - 4 a^{(3)}_1 + \frac58 \Bigr] \frac{(N_e - 2) (N_e - 3) (N_e - 4) (N_e - 5)}{24} \nonumber \\
 a^{(N_e=6)}_0 &=& a^{(2)}_0 + \Bigl[ a^{(3)}_0 - a^{(2)}_0 \Bigr] (N_e - 2) + \Bigl[ a^{(4)}_0 - 2 a^{(3)}_0 + a^{(2)}_0  \Bigr] \frac{(N_e - 2) (N_e - 3)}{2} \nonumber \\
 &+& \Bigl[ a^{(5)}_2 - 3 a^{(4)}_2 + 3 a^{(3)}_2 - a^{(2)}_2 \Bigr] \frac{(N_e - 2) (N_e - 3) (N_e - 4)}{6} \; \; \; \label{coefa3d} \\
 &+& \Bigl[ a^{(6)}_2 - 4 a^{(5)}_2 + 6 a^{(4)}_2 - 4 a^{(3)}_2 + a^{(2)}_2 \Bigr] \frac{(N_e - 2) (N_e - 3) (N_e - 4) (N_e - 5)}{24} \nonumber
\end{eqnarray}
and
\begin{eqnarray}
 b^{(N_e=6)}_k &=& b^{(2)}_k + \Bigl[ b^{(3)}_k - b^{(2)}_k \Bigr] (N_e - 2) + \Bigl[ b^{(4)}_k - 2 b^{(3)}_k + b^{(2)}_k \Bigr] \frac{(N_e - 2) (N_e - 3)}{2} \nonumber \\
 &+& \Bigl[ b^{(5)}_k - 3 b^{(4)}_k + 3 b^{(3)}_k - b^{(2)}_k \Bigr] \frac{(N_e - 2) (N_e - 3) (N_e - 4)}{6} \; \; \; \label{coefb3d} \\
 &+& \Bigl[ b^{(6)}_k - 4 b^{(5)}_k + 6 b^{(4)}_k - 4 b^{(3)}_k + b^{(2)}_k \Bigr] \frac{(N_e - 2) (N_e - 3) (N_e - 4) (N_e - 5)}{24} \nonumber
\end{eqnarray}
where in the right-hand sides of these equations we must put $N_e = 6$. Generalization of these formulas to the cases when $N_e \ge 7$ is straightforward and relatively simple.

As follows from these formulas the addition of one electron, i.e. $N_e - 1 \rightarrow N_e$, affects only the last term in each of these formulas, while all previous terms contains coefficients known for 
atomic systems with $N_e - 1$ electrons. This allows one to consider different atomic $LS$-terms for the bound states with different $N_e$ (it is assumed that the `selection rules' mentioned in the main 
text are obeyed for these $LS$-terms).

\newpage
 \begin{table}[tbp]
   \caption{The total non-relativistic energies $E$ of the different atoms/ions in their ground states in atomic units. 
            All nuclear masses are infinite. $Q$ is the nuclear electric charge and $N_e$ is the total number of bounded
            electrons.}
     \begin{center}
     \begin{tabular}{| c | c | c | c |}
      \hline\hline
  $Q$ &             $N_e = 2$                &     $N_e = 3$       &     $N_e = 4$    \\
     \hline\hline
   1 &   -0.5277510165443771965925           &  -----------------  &  ---------------- \\      
   2 &   -2.90372437703411959831115924519440 &  -----------------  &  ---------------- \\  
   3 &   -7.27991341266930596491875          &   -7.4780603236503  &  ---------------- \\  
   4 &  -13.65556623842358670208051          &  -14.3247631764654  &  -14.667356407951 \\
   5 &  -22.03097158024278154165469          &  -23.424605720957   &  -24.348884381902 \\
   6 &  -32.40624660189853031055685          &  -34.775511275626   &  -36.534852285202 \\
             \hline 
   7 &  -44.781445148772704645183         &  -48.376898319137   &  -51.222712616143 \\
   8 &  -59.156595122757925558542         &  -64.228542082701   &  -68.411541657589 \\
   9 &  -75.531712363959491104856         &  -82.330338097298   &  -88.100927676354 \\
  10 &  -93.906806515037549421417         & -102.682231482398   & -110.290661070069 \\
  11 & -114.28188377607272189582          & -125.2841907536473  & -134.980624604257 \\
  12 & -136.65694831264692990427          & -150.1361966044594  & -162.170747906692 \\
             \hline 
  13 & -161.03200302605835987252          & -177.238236559961   & -191.860986338262 \\
  14 & -187.40704999866292631487          & -206.5903022122780  & -224.051310298012 \\
  15 & -215.78209076353716023462          & -238.1923876941461  & -258.741699427160 \\
  16 & -246.15712647425473932009          & -272.0444887900725  & -295.932139288646 \\
  17 & -278.53215801540009570337          & -308.1466023952556  & -335.622619375075 \\
  18 & -312.90718607661114879880          & -346.4987261736714  & -377.813131866050 \\
             \hline 
  19 & -349.28221120345316700447          & -387.1008583345610  & -422.503670826658 \\
  20 & -387.65723383315855621790          & -429.9529974827626  & -469.694231675265 \\
  21 & -428.03225432023469116264          & -475.0551425155     & -519.384810821074 \\
  22 & -470.40727295513838395930          & -522.4072925498     & -571.575405411671 \\
  23 & -514.78228997811177388135          & -572.0094468708     & -626.266013153662 \\
  24 & -561.15730558958127234352          & -623.8616048933     & -683.456632182920 \\
  25 & -609.53231995807574620568          & -677.9637661344     & -743.147260969064 \\
              \hline
  26 & -659.90733322632780520901          & -734.3159301916     & -805.337898245040 \\
  27 & -712.28234551602655145614          & -792.9180967274     & -870.028542951686 \\
  28 & -766.65735693155709991040          & -853.7702654564     & -937.219194199135 \\
  30 & ------------------------           &  -----------------  & -1079.100513407098 \\
  36 & ------------------------           &  -----------------  & -1564.744568198454 \\
     \hline\hline
  \end{tabular}
  \end{center}
  \end{table}

\begin{table}[tbp]
   \caption{Coefficients $a^{(N_e)}_2, a^{(N_e)}_1, a^{(N_e)}_0$ and $b^{(N_e)}_1, b^{(N_e)}_2, \ldots, b^{(N_e)}_7$
            in the ten-term expansion of the $E(Q,N_e)-$function (see, Eq.(12)).}
     \begin{center}
     \begin{tabular}{| c | c | c | c |}
      \hline\hline
  coefficients   &  $N_e = 2$     &   $N_e = 3$    &   $N_e = 4$    \\
     \hline\hline
  $a^{(N_e)}_2$  &  -1.0000000000 &  -1.1249999945 &  -1.2499999978 \\

  $a^{(N_e)}_1$  &   0.6249999961 &   1.0228047058 &   1.5592739061 \\

  $a^{(N_e)}_0$  &  -0.1576662873 &  -0.4081459572 &  -0.8771019520 \\
              \hline
  $b^{(N_e)}_1$  &   0.0086962674 &  -0.0170061678 &  -0.0429178704 \\

  $b^{(N_e)}_2$  &  -0.0008567981 &  -0.0341594959 &  -0.1694257553 \\

  $b^{(N_e)}_3$  &  -0.0012640360 &  -0.1084597316 &  -0.3362058744 \\

  $b^{(N_e)}_4$  &   0.0003957668 &   0.2941494433 &   0.6932000874 \\

  $b^{(N_e)}_5$  &  -0.0030653579 &  -1.4458330710 &  -5.7572871897 \\

  $b^{(N_e)}_6$  &   0.0037323637 &   2.9184336113 &  14.3319856750 \\

  $b^{(N_e)}_7$  &  -0.0027229312 &  -3.3508481226 & -22.3281869042 \\
     \hline\hline
  \end{tabular}
  \end{center}
  \end{table}


\begin{thebibliography}{10}
\bibitem{TomFer} This problem has a unique solution for the Tomas-Fermi method, where one finds that $T = -\frac{3}{7} U_{en}$ and $U_{ee} = -\frac{1}{7} U_{en}$. This 
means that the total energy is $E = \frac{3}{7} U_{en}$, where $T, U_{en}, U_{ee}$ are the expectation values of the (electron) kinetic energy, electron-nucleus 
attraction and electron-electron repulsion, respectively.
\end{thebibliography}
\end{document}